\documentclass{article}
\usepackage[preprint]{spconf}
\usepackage{amsmath}
\usepackage{hyperref}
\usepackage{tikz}
\usetikzlibrary{calc,arrows,shapes,positioning,automata,decorations.pathmorphing}
\usepackage{tkz-euclide}
\usepackage{float}
\usepackage{array}
\usepackage{bm}
\usepackage{siunitx}  

\begin{document}
\ninept


\tikzset{
block/.style = {draw, fill=white, rectangle, minimum height=3em, minimum width=3em},
tmp/.style  = {coordinate}, 
circ/.style= {draw, fill=white, circle, node distance=1cm, minimum size=0.6cm},
input/.style = {coordinate},
output/.style= {coordinate},
pinstyle/.style = {pin edge={to-,thin,black}.},
snake it/.style={decorate, decoration=snake}
}

\title{RADE for Land Mobile Radio: A Neural Codec for Transmission of Speech over Baseband FM Radio Channels}
\twoauthors                                                                                             
 {David Rowe \sthanks{Supported by a grant from Amateur Radio Digital Communications}}                             
       {Rowetel P/L \\
        Adelaide, South Australia}                                                                                   
 {Tibor Bece}                                                                                   
       {Tibicom P/L \\
        Goulburn, New South Wales}                                                                                 
 
\maketitle

\begin{abstract}
In the 1990s Land Mobile Radio (LMR) systems evolved from analog frequency modulation (FM) to standardised digital systems. 
Both digital and analog FM systems now co-exist in various services and exhibit similar speech quality. 
The architecture of many digital radios retains the analog FM modulator and demodulator from legacy analog radios, but driven by a multi-level digital pulse train rather than an analog voice signal. 
We denote this architecture baseband FM (BBFM). 
In this paper we describe a modern machine learning approach that uses an autoencoder to send high quality, 8 kHz bandwidth speech over the BBFM channel. 
The speech quality is shown to be superior to analog FM over simulated LMR channels in the presence of fading, and a demonstration of the system running over commodity UHF radios is presented.
\end{abstract}

\section{Introduction}

Many land mobile radio (LMR) systems \cite{kunavut2014overview} for push to talk (PTT) VHF/UHF voice are built on radio hardware employing the baseband frequency modulation (BBFM) architecture.
Applications for LMR include public safety (police, fire, ambulance services), commercial (taxis, mining industry, transport), and personal (off road recreational vehicles, amateur radio).
Analog frequency modulation (FM) remains popular (e.g. millions of radios in Australia alone) because of its speech quality, low cost, interoperability and ability to operate without infrastructure.
Key requirements for LMR are speech quality, spectral efficiency, and robustness to fading due to vehicle or hand held radio movement. 
Weak signal/low SNR performance is a secondary requirement.

The ubiquitous BBFM radio architecture still enjoys a great deal of popularity over more modern SDR designs for several reasons:
\begin{enumerate}
\item It is inexpensive, a complete handset may retail for \$USD40.
\item Insensitive to frequency and phase offsets, frequency accuracy, and modulation index. 
Low cost frequency references can be used.
\item Transmitter compliance with spurious emissions requirements is easier than SDRs as there is no carrier feed through, or DAC and mixer non-linearities. 
\item Only coarse (frame sync) and fine timing estimation is required - no phase or frequency estimation, making demodulator implementation straight forward and robust.
\item The waveform can be also generated and received by modern IQ SDR radios, so offers a degree of future proofing.
\end{enumerate}

The speech quality of digital LMR systems employing the BBFM architecture is close to analog FM \cite{atkinson2012intelligibility}\cite{vanderau1998delivered}; has not evolved since the mid 1990s; and is constrained to an audio bandwidth of less than 4~kHz.
Alternatives for higher speech quality include enhancements to LTE 4G communications to support public safety applications \cite{jarwan2019lte} such as mission critical PTT (MCPTT).
However these systems require handsets with a parallel, non BBFM radio architecture, and cannot operate in remote areas without LTE coverage, or in simplex (handset to handset) mode.

This paper describes a radio autoencoder \cite{oshea2017introductiondeeplearningphysical} derived from the RDO-VAE structure of DRED \cite{valin2024dred}. 
It sends 8~kHz bandwidth speech over the same BBFM architecture as existing analog and digital LMR systems, but with significant improved speech quality and robustness to fading. 
Borrowing from our HF work \cite{rowe2025radeneuralcodectransmitting}, we denote the machine learning parts of the system RADE (RADio autoEncoder).
Other approaches using machine learning (ML) to send speech over radio channels include the very low latency work of \cite{bokaei2025low}.  Our work differs in the focus on the BBFM channel for the LMR use case, and the use of a vocoder feature set rather than direct encoding of speech.

\begin{figure}[h]
\begin{center}
\begin{tikzpicture}[auto, node distance=1.75cm,>=triangle 45,x=1.0cm,y=1.0cm,align=center,text width=1.15cm,font=\footnotesize]

\node [input] (rinput) {};
\node [block, right of=rinput, node distance=1.25cm] (speech_enc) {Speech Enc};
\node [block, right of=speech_enc] (fec_enc) {FEC Enc};
\node [block, right of=fec_enc] (tx_filt) {Tx Filter};
\node [block, right of=tx_filt] (fm_mod) {FM Mod};
\node [block, below of=fm_mod,node distance=2cm] (fm_demod) {FM Demod};
\node [block, left of=fm_demod] (rx_filt) {Rx Filter};
\node [block, left of=rx_filt] (fec_dec) {FEC Decoder};
\node [block, left of=fec_dec] (speech_dec) {Speech Dec};
\node [output, left of=speech_dec,node distance=1.25cm] (routput) {};

\draw [->] node[above,text width=1cm] {Input Speech} (rinput) -- (speech_enc);
\draw [->] (speech_enc) -- (fec_enc);
\draw [->] (fec_enc) -- (tx_filt);
\draw [->] (tx_filt) -- (fm_mod);
\path [draw=blue, snake it] (fm_mod) -- node[left] {Radio Channel} (fm_demod);
\draw [->] (fm_demod) -- (rx_filt);
\draw [->] (rx_filt) -- (fec_dec);
\draw [->] (fec_dec) -- (speech_dec);
\draw [->] (speech_dec) -- (routput) node[above, text width=1cm] {Output Speech};

\end{tikzpicture}
\end{center}
\caption{Classical DSP Speech over BBFM System}
\label{fig:classical_block}
\end{figure}
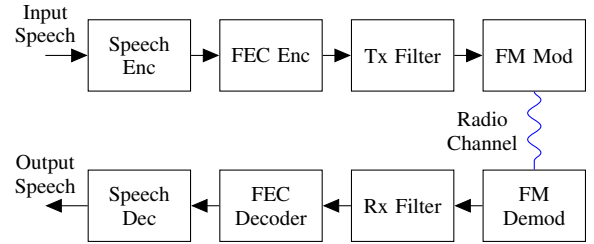

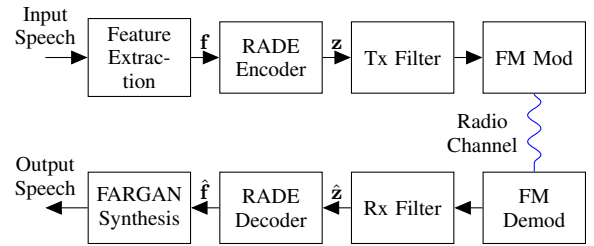
\begin{figure}[h]
\begin{center}
\begin{tikzpicture}[auto, node distance=1.75cm,>=triangle 45,x=1.0cm,y=1.0cm,align=center,text width=1.15cm, font=\footnotesize]

\node [input] (rinput) {};
\node [block, right of=rinput, node distance=1.25cm] (feature_ext) {Feature Extraction};
\node [block, right of=feature_ext] (rade_enc) {RADE Encoder};
\node [block, right of=rade_enc] (tx_filt) {Tx Filter};
\node [block, right of=tx_filt] (fm_mod) {FM Mod};
\node [block, below of=fm_mod,node distance=2cm] (fm_demod) {FM Demod};
\node [block, left of=fm_demod] (rx_filt) {Rx Filter};
\node [block, left of=rx_filt] (rade_dec) {RADE Decoder};
\node [block, left of=rade_dec] (fargan) {FARGAN Synthesis};
\node [output, left of=fargan,node distance=1.25cm] (routput) {};

\draw [->] node[above,text width=1cm] {Input Speech} (rinput) -- (feature_ext);
\draw [->] (feature_ext) -- node[above, text width=0.5cm] {$\bf{f}$} (rade_enc);
\draw [->] (rade_enc) -- node[above, text width=0.5cm] {$\bf{z}$} (tx_filt);
\draw [->] (tx_filt) -- (fm_mod);
\path [draw=blue, snake it] (fm_mod) -- node[left] {Radio Channel} (fm_demod);
\draw [->] (fm_demod) -- (rx_filt);
\draw [->] (rx_filt) -- node[above, text width=0.5cm] {$\hat{\bf{z}}$} (rade_dec);
\draw [->] (rade_dec) -- node[above, text width=0.5cm] {$\hat{\bf{f}}$} (fargan);
\draw [->] (fargan) -- (routput) node[above, text width=1cm] {Output Speech};

\end{tikzpicture}
\end{center}
\caption{Radio Autoencoder over BBFM System}
\label{fig:rade_block}
\end{figure}

Consider the land mobile radio (LMR) system based on classical DSP shown in Figure \ref{fig:classical_block}. 
Speech with a bandwidth of around 3~kHz is sampled at 8~kHz and compressed to a low bit rate using a speech encoder. 
Forward error correction (FEC) adds redundant bits to protect the sensitive payload speech bits from channel errors. 
Framing (not illustrated) concatenates several speech encoder frames and adds frame sync and signalling information such as station ID. 
Payload bits are mapped to a discretely valued (2 or 4 level) sequence of analog shift keyed (ASK) symbols driving an analog FM modulator. 
The symbols may be filtered to minimise inter-symbol interference (ISI) and constrain the spectrum of the transmitted signal. 
The receiver employs a limiter and analog FM demodulator to recover the symbols, which are converted to bits with a slicer. 
The FEC decoder attempts to correct any bit errors, and the speech decoder converts the signal back into a sampled speech signal for replay over a loudspeaker.

Figure \ref{fig:rade_block} presents a radio autoencoder for the BBFM channel. 
A feature extractor takes  speech sampled at 16~kHz and generates a set of vocoder feature vectors (short term spectrum, pitch, voicing) $\mathbf{f}$. 
The RADE encoder uses a neural network to directly generate a vector of ASK symbols $\mathbf{z}$ from the feature vectors $\mathbf{f}$.
These ASK symbols are sent over the BBFM channel to the RADE decoder which transforms them back to vocoder features $\hat{\mathbf{f}}$.
We employ the FARGAN vocoder \cite{valin2024low} for high quality speech synthesis with an audio bandwidth of 8~kHz.

\begin{figure}
\begin{center}
\scalebox{0.7}{\input {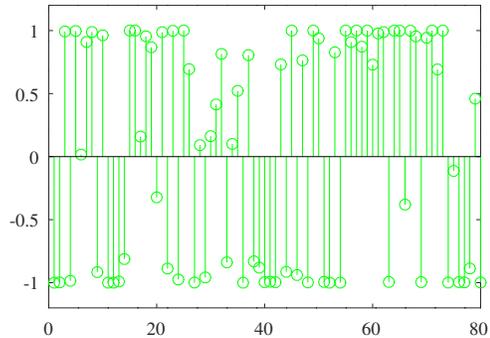}}
\end{center}
\caption{$d=80$ ASK symbols from RADE encoder representing 40~ms of encoded speech. 
The encoder symbols are continuously valued and not part of a discrete constellation.}
\label{fig:ask_scatter}
\end{figure}

Our contributions are:
\begin{enumerate}
\item A neural network (radio autoencoder) for sending speech over the BBFM channel that significantly outperforms classical analog and digital alternatives.
\item A neural network for sending high quality, 8~kHz bandwidth speech over the BBFM channel, more than double the audio bandwidth of classical analog and digital alternatives.
\item An autoencoder that combines the classical DSP functions of quantisation and channel coding to generate discrete time but continuously valued (analog) ASK symbols directly from vocoder features. 
Unlike classical approaches there is no intermediate bit stream, and the ASK symbols (Figure \ref{fig:ask_scatter}) emerge from the training process rather than being members of a well defined, discrete constellation.
\item A linear approximation of the BBFM channel to support training and simulation of ML networks.
\end{enumerate}

The design of the ML components is presented in Section \ref{sec:design}. 
The FM modulation and demodulation process is highly non-linear. 
In Section \ref{sec:bbfm_model} we develop a linearised model for ASK pulses over BBFM when the radio signal is subject to AWGN and fading channel impairments. 
We then describe how this model is used to train the radio autoencoder in Section \ref{sec:training}. 
In Section \ref{sec:asr_results} we use automatic speech recognition (ASR) to evaluate the RADE encoded BBFM over AWGN and fading channels, and compare results to analog FM. 
In Section \ref{sec:demo} we describe a demonstration system running over radio hardware and provide links to speech samples from our system, analog FM, and digital LMR systems for comparison.

\section{Design}
\label{sec:design}

RADE uses 20-dimensional feature vectors $\mathbf{f}$ that consist of 18~Bark-scale cepstral coefficients, the pitch period, and a voicing parameter ~\cite{valin2019lpcnet}.
The RADE encoder and decoder form an autoencoder that is trained to minimise the reconstruction loss $\mathcal{L}(\mathbf{f}, \hat{\mathbf{f}})$ between the input features~$\mathbf{f}$ and the decoded features~$\hat{\mathbf{f}}$, as defined in Eq.~(12) of~\cite{valin2024dred}.
The feature vectors $\mathbf{f}$ are generated every 10~ms, concatenated, and passed to the RADE encoder every 40~ms (frame rate of 25~Hz).
They are transformed to a $d=80$ dimensional ASK symbol vector $\mathbf{z}$ by a stack of 1D convolutional (conv) and gated recurrent units (GRU), arranged in a DenseNet-like~\cite{huang2017densely} topology.
The RADE encoder and decoder are the same as employed for our HF work; further details are provided in \cite{rowe2025radeneuralcodectransmitting}.

The feature vectors $\mathbf{f}$ are updated at 100~Hz, and the dimension $d=80$ ASK symbol vectors $\mathbf{z}$ at 25~Hz, i.e. a symbol rate of $R_s=2000$ symb/s for the payload speech information. 
The resources for the ML processing are around 1 Mbyte of read only storage and 32 million multiply-accumulates (MMACs) each for the encoder and decoder, with overall CPU load being dominated by the FARGAN vocoder (300 MMACS) \cite{rowe2025radeneuralcodectransmitting}.

Our system is not sensitive to the design of the pulse shaping filters, apart from the assumption that they minimise inter-symbol interference (ISI). 
Not shown in Figure \ref{fig:rade_block} are other components such as the DAC/ADC, RF frequency translation, IF filtering, limiter and power amplifier. 
All of these components and the pulse shaping filters are assumed to have a unity transfer function with respect to the ASK pulses and minimal impact on system performance.
As is common with other digital-over-BBFM systems, no de-emphasis or pre-emphasis is employed.
All blocks not marked RADE are implemented with classical analog or digital signal processing, common to existing analog or digital FM radios, and requiring modest amounts of CPU and memory compared to the machine learning (ML) components.

\section{BBFM Channel Model}
\label{sec:bbfm_model}

In this section we develop a linearised model of the BBFM signal processing in AWGN and multipath fading channels to support simulation and training.
We assume an ideal FM receiver that is sensitive only to frequency modulation of the received signal - a reasonable assumption in practice due to the availability of low cost FM demodulators with good performance.

Above threshold, the signal to noise ratio (SNR) at the output of the FM demodulator as a function of the input carrier to noise ratio (CNR) is given by \cite{crilly2009communication}:
\begin{equation}
\label{eq:carlson}
\mathrm{SNR} = 3 \beta^2 \overline{x^2} \mathrm{CNR}
\end{equation}
where $\beta=f_d / f_m$ is the modulation index, $f_d$ is the maximum deviation, $f_m$ is the maximum frequency of the modulation signal $x(t)$, $\overline{x^2}$ is the mean power of $x(t)$, and $x(t)$ is constrained to have a peak of 1 which corresponds to the maximum deviation $f_d$. 
For sinusoidal modulation (common for testing) where $x(t)=A\mathrm{cos}(\omega_m t)$, $\overline{x^2}=A^2/2$. 
Practical FM receiver measurements are often performed using sinusoidal $x(t)$ with $A<1$, e.g. 
$A=0.6$. 
Equation (\ref{eq:carlson}) holds above a threshold SNR which is typically around 12~dB at the FM demodulator output.

The demodulator input CNR is:
\begin{equation}
\mathrm{CNR}=\frac{C}{N_0f_m}
\end{equation}
where $C$ is the carrier power, and $N_0$ the spectral noise density. 
Note the noise power is measured in the bandwidth of the modulating signal $f_m$. 
The demodulator output SNR is also measured in a noise bandwidth of $f_m$. 
The somewhat counter-intuitive convention of measuring input CNR in the bandwidth of the modulating signal (rather than the modulated FM signal) allows a simple comparison to linear modulation methods such as AM and SSB \cite{crilly2009communication}.

In this paper, we will express SNR as a function of the received signal power R in milliwatts. 
 Expressing the demodulator input CNR as a function of R, and defining the spectral noise density as the thermal noise $N_0=kT$ we obtain:
\begin{equation}
\label{eq:cnr_r}
\mathrm{CNR} = \frac{R}{10^3kT F f_m} 
\end{equation}
where $k$ is the Boltzmann constant, and $T=274$~K, and $F$ is the noise factor of the radio. 
 Substituting into (\ref{eq:carlson}):
\begin{equation}
\begin{split}
\mathrm{SNR} &= \frac{3 \beta^2 \overline{x^2} R}{10^3kT F f_m} \\
\mathrm{SNR_{dB}} &= \mathrm{R_{dBm}} + \mathrm{G_{FM}} \\
\mathrm{G_{FM}} &= 10\mathrm{log_{10}}\left(\frac{3 \beta^2 \overline{x^2}}{10^3kTf_m} \right) - \mathrm{NF_{dB}}
\end{split}
\end{equation}
where the noise factor F has been expressed in the more convenient noise figure $\mathrm{NF_{dB}}=10\mathrm{log_{10}}(F)$. 
Using some typical values for a LMR receiver as $\mathrm{NF_{dB}}=5$ dB, $f_d=2.5$~kHz, $f_m=3$~kHz, and a sinusoidal modulation signal with $A=1,\overline{x^2}=0.5$, we obtain:
\begin{equation}
\mathrm{SNR_{dB}} = \mathrm{R_{dBm}} + 134.41
\end{equation}
For example at the threshold $\mathrm{SNR_{dB}}=12$~dB, $\mathrm{R_{dBm}}=-122.41$~dBm.

Our previous experiments show the HF RADE system \cite{rowe2025radeneuralcodectransmitting} can provide high quality speech at around 0~dB SNR, suggesting we have ample link margin and should obtain high quality speech over the BBFM channel down to the threshold SNR.

Consider a typical LMR channel from a base station transmitter to a receiver that is pedestrian or vehicle mobile. 
The propagation path will typically not be line of site, but will reflect off several terrestrial objects (such as buildings) to reach the receiver. 
The relative phase shifts of the paths will result in multipath fading, which due to the receiver movement will be time varying. 
The multipath channel typically evolves slowly (e.g. 
a bandwidth of a few 10's of Hz at 60 km/hr) compared to the symbol rate (kHz), so the rate of change in phase has only a trivial impact on the output of the FM demodulator compared to the data symbols and can be ignored. 
We can therefore model the multipath channel as Rayleigh distributed magnitude fading $|H|$ and AWGN noise added to the FM carrier. 
We have chosen the two path fading model described in the faded channel simulator section of \cite{telecommunications2016digital}:
\begin{equation}
y(t) = x(t)G_1(t) + x(t-d)G_2(t)
\end{equation}
where $x(t)$ is the signal from the transmitter, and $y(t)$ is the output of the multipath fading model.
$G_1$ and $G_2$ are two band-limited complex Gaussian signals with \emph{Doppler Spread} bandwidth $B_{ds}$, and $d$ is the delay spread (path delay) in seconds. 
$B_{ds}$ is derived from the vehicle speed as:
\begin{equation}
B_{ds} = 2f_c v/3\times10^8
\end{equation}
where $f_c$ is the carrier frequency and $v$ is the vehicle speed in meters per second.
By taking the $z$-transform $|H|$ can be computed as:
\begin{equation}
\label{eq:fading mag}
|H| = |G_1+e^{-j d F_s}G_2|
\end{equation}
$G_1$ and $G_2$ are time varying and updated at a suitable sample rate.

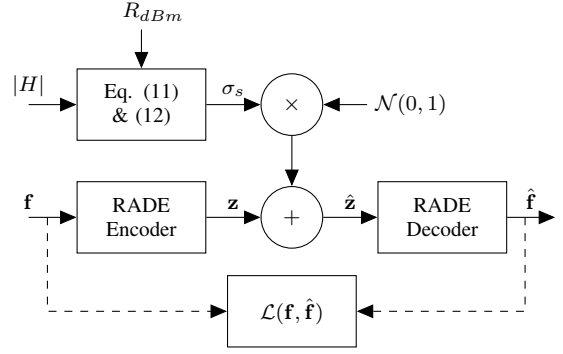
\begin{figure}
\begin{center}
\begin{tikzpicture}[auto, node distance=2cm,>=triangle 45,x=1.0cm,y=1.0cm,align=center,text width=1.5cm, font=\footnotesize]

\node [input] (rinput) {};
\node [tmp, right of=rinput,node distance=0.25cm] (rinput1) {};
\node [block, right of=rinput,node distance=1.5cm] (rade_enc) {RADE Encoder};
\node [circ, right of=rade_enc, node distance=2cm, text width=0.5cm] (add) {$+$};
\node [circ, above of=add, node distance=1.5cm, text width=0.5cm] (mult) {$\times$};
\node [block, left of=mult] (lin_model) {Eq. (\ref{eq:fm_snr}) \& (\ref{eq:fm_sigma})};
\node [block, right of=add] (rade_dec) {RADE Decoder};
\node [output, right of=rade_dec,node distance=1.5cm] (routput) {};
\node [tmp, left of=routput,node distance=0.4cm] (routput1) {};
\node [block, below of=add, node distance=1.25cm] (loss_fn) {$\mathcal{L}(\mathbf{f}, \hat{\mathbf{f}})$};

\draw [->] node[above,text width=1cm] {$\bf{f}$} (rinput) -- (rade_enc);
\draw [->] (rade_enc) -- node[above, text width=0.5cm] {$\bf{z}$} (add);
\draw [->] (mult) -- (add);
\draw [->] (add) -- node[above, text width=0.5cm] {$\hat{\bf{z}}$} (rade_dec);
\draw [->] (rade_dec) -- node[above, text width=1cm] {$\hat{\bf{f}}$} (routput);

\node [input, above of=lin_model,node distance=1cm] (setpoint) {};
\node [input, left of=lin_model,node distance=1.5cm] (fading) {};
\node [input, right of=mult,node distance=1cm] (awgn) {};
\draw [->] (setpoint) node[above, text width=0.5cm] {$R_{dBm}$} -- (lin_model); 
\draw [->] (fading) node[above, text width=0.5cm] {$|H|$} -- (lin_model); 
\draw [->] (lin_model) -- node[above, text width=0.5cm] {$\sigma_s$} (mult);
\draw [->] (awgn) node[right, text width=0.75cm] {$\mathcal{N}(0,1)$} -- (mult); 
\draw [dashed,->] (rinput1) |- (loss_fn); 
\draw [dashed,->] (routput1)|- (loss_fn); 

\end{tikzpicture}
\end{center}
\caption{Linearised BBFM model. 
The fading $|H|$ modulates the noise about the set point defined by $R_{dBm}$ via Eq. (\ref{eq:fm_snr}) \& (\ref{eq:fm_sigma}); the magnitude of the symbols $\bf{z}$ is set by the deviation $f_d$ and remains constant. The loss function $\mathcal{L}(\mathbf{f}, \hat{\mathbf{f}})$ is used during training to minimise end to end distortion of the vocoder features.}
\label{fig:bbfm_linear}
\end{figure}

To simplify simulation and training we model the filtered received symbols from the FM demodulator as the transmitted symbols with additive Gaussian noise:
\begin{equation}
\label{eq:fm_noise}
\hat{\bf{z}} = \bf{z} + \bf{n}
\end{equation}
where $\bf{n}$ is a vector of noise samples such that the power of each noise sample is time varying: 
\begin{equation}
n_i = \sigma_s^i\mathcal{N}(0,1) 
\end{equation}
Note this approximates the triangular spectra (noise power increasing with frequency) of FM noise \cite[p.~457]{crilly2009communication} with flat noise of the same power. 
In the following we omit the sample index $i$ for clarity such that $\sigma_s=\sigma_s^i$. 
The mean magnitude and hence power of each symbol in $\hat{\bf{z}}$ at the output of a FM demodulator is a function of the deviation and not the channel CNR. 
Thus as the CNR is modulated by fading, the mean value of each symbol in $\hat{\bf{z}}$ will be constant, however the FM demodulator output SNR will change, implying that output noise power $\sigma_s^2$ is modulated as a function of $|H|$ (Figure \ref{fig:bbfm_linear}).

Fading may push the input CNR beneath the demodulator threshold. 
At this point the SNR drops rapidly, and the noise becomes impulsive rather than AWGN. 
Except for rare deep fades, our system is designed to operate above threshold. 
For simplicity of training, we approximate the demodulator SNR using a piecewise linear model where SNR drops rapidly beneath a received power threshold $T_{dBm}$ corresponding to a demodulator output SNR of 12~dB:

\begin{equation}
\begin{split}
\label{eq:fm_snr}
\mathrm{SNR_{dB}} &= \begin{cases}
           \mathrm{R_{dBm}^\prime} + \mathrm{G_{FM}}, & \mathrm{R_{dBm}^\prime} \ge \mathrm{T_{dBm}} \\
           3\mathrm{R_{dBm}^\prime} + \mathrm{G_{FM}} - 2\mathrm{T_{dBm}}, & \mathrm{R_{dBm}^\prime} < \mathrm{T_{dBm}}
	       \end{cases} \\
\mathrm{R_{dBm}^\prime} &= \mathrm{R_{dBm}}+\mathrm{H_{dB}}
\end{split}
\end{equation}
where $\mathrm{R_{dBm}}$ is the mean (set point) received power, and $\mathrm{H_{dB}}=20\mathrm{log_{10}}|H|$ is the current magnitude of the fading channel, and $\mathrm{T_{dBm}}=12-\mathrm{G_{FM}}$.
In the sub-threshold region (second clause of (\ref{eq:fm_snr})) a gradient of 3 was selected.
As the theoretical performance of the FM demodulator in the sub-threshold 
region is highly non-linear and signal dependent, the gradient of 3
was selected as a good approximation to real, measured systems.

\begin{figure}
\begin{center}
\scalebox{.7}{\input {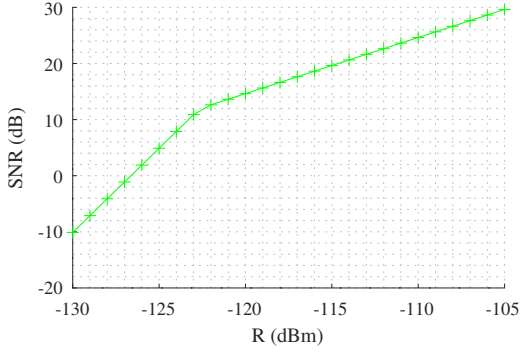}}
\end{center}
\caption{Piecewise SNR versus $R_{dBm}$ for FM demodulator model in (\ref{eq:fm_snr}) with $\beta=2.5/3$, a sinusoidal modulation signal $x(t)$ with $A=1$, $NF_{dB}=5$ dB, for an AWGN channel ($H=1$).}
\label{fig:fm_r_snr}
\end{figure}

Figure \ref{fig:fm_r_snr} plots (\ref{eq:fm_snr}) for an AWGN channel ($H=1$).

For simulation of the RADE system over the BBFM channel, we approximate the combined effect of the pulse shaping Tx and Rx filters as simple band limiting to the symbol rate $R_s$. 
Table \ref{tab:snr_model_params} contains the model parameters we have used for analog FM and RADE simulation over the BBFM channel model.

To calculate the noise power to apply in simulation we solve for $\sigma_s$:
\begin{equation}
\label{eq:fm_sigma}
\begin{split}
\frac{S}{\sigma_s^2} &= \mathrm{SNR} \\
\sigma_s &= \frac{A}{\sqrt{\mathrm{SNR}}}
\end{split}
\end{equation}
where $A$ is the amplitude of the symbol corresponding to maximum deviation $f_d$. 
 As $|H|$ evolves over time, $\sigma_s$ should be re-calculated for every symbol using the latest sample of $|H|$. 
The channel simulation procedure is:
\begin{enumerate}
\item Select a set point (mean) $\mathrm{R_{dBm}}$ and vehicle speed.
\item Generate a set of $|H|$ samples at the symbol rate $R_s$ based on a vehicle speed from the fading channel simulator  (\ref{eq:fading mag}).
\item Calculate SNR for each $|H|$ sample using (\ref{eq:fm_snr}). 

\item Set $\sigma_s$ for each symbol in the set using (\ref{eq:fm_sigma}).
\item Apply noise to each symbol using (\ref{eq:fm_noise}).
\end{enumerate}

\begin{table}
\caption{Channel Model parameters}
\label{tab:snr_model_params}
\centering
\begin{tabular}{l c c}
\hline
Parameter & Analog FM & RADE \\
\hline
Carrier frequency $f_c$ & 450 MHz & 450 MHz \\ 
Symbol rate $R_s$ & - & 2000 \\
Deviation $f_d$ & 2500 & 1800 \\
Maximum $x(t)$ freq. 
$f_m$ & 3000 & 2880 \\
Noise figure $\mathrm{NF_{dB}}$ & 5 & 5 \\
\hline
\end{tabular}
\end{table}

To support comparative testing, analog FM samples can be simulated with the same linearised model  as the RADE system, with the noise scaled such that the SNR was referenced to a noise bandwidth $f_m$~Hz. 
Speech is band pass filtered and limited to simulate the signal processing in a typical commercial FM radio.
Typical peak to average power ratio (PAPR) of the speech tested here is 15~dB before compression and 8~dB after compression.

\begin{figure}
\begin{center}
\begin{tikzpicture}[auto, node distance=2cm,>=triangle 45,x=1.0cm,y=1.0cm,align=center,text width=1.25cm,font=\footnotesize]

\node [input] (rinput) {};
\node [block, right of=rinput, node distance=1.25cm] (bpf1) {BPF 300-3000};
\node [block, right of=bpf1] (pre) {Pre-emphasis};
\node [block, right of=pre,text width=1.5cm] (hilb_comp) {Limiter};
\node [block, right of=hilb_comp] (bpf2) {BPF 300-3000};
\node [block, below of=bpf2,node distance=2.5cm] (de) {De-emphasis};
\node [circ,left of=de,text width=0.5cm,node distance=2cm] (add) {$+$};
\node [block, left of=add] (bpf3) {BPF 300-3000};
\node [output, left of=bpf3,node distance=1.25cm] (routput) {};

\draw [->] node[above,text width=1cm] {Input Speech} (rinput) -- (bpf1);
\draw [->] (bpf1) -- (pre);
\draw [->] (pre) -- (hilb_comp);
\draw [->] (hilb_comp) -- (bpf2);
\draw [->] (bpf2) -- (de);
\draw [->] (de) -- (add);
\draw [->] (add) -- (bpf3);
\draw [->] (bpf3) -- (routput) node[above, text width=1cm] {Output Speech};

\node [input, above of=add,node distance=1cm] (awgn) {};
\draw [->] (awgn) node[above, text width=0.75cm] {$\mathcal{N}(0,\sigma_s^2)$} -- (add); 

\end{tikzpicture}
\end{center}
\caption{Analog FM simulation. The linearised BBFM model from Eq. (\ref{eq:fm_snr}) \& (\ref{eq:fm_sigma}) is used to obtain $\sigma_s$.}
\label{fig:fm_simulation}
\end{figure}
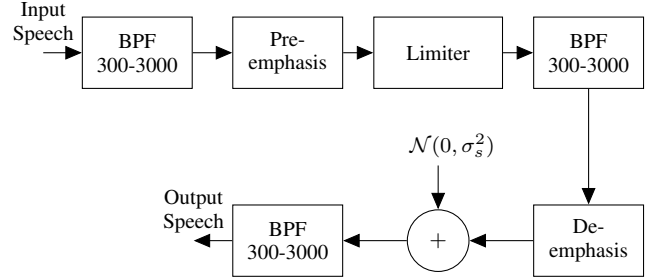

For a given $R_{dBm}$, Eq. (\ref{eq:fm_snr}) provides the demodulator output SNR.
To calculate $G_{fm}$, we use $A=1, \overline{x^2}=0.5$.
The SNR given by Eq. (\ref{eq:fm_snr}) has a noise power measured in a bandwidth of $f_m$; the simulated noise is generated at $F_s=8$~kHz with the same noise density.
Speech samples are gain controlled so the peak level just reaches $A=1$, which results in a measured $\overline{x^2}=0.07$ for speech signals, around 8~dB less (equivalent to the speech PAPR) than the $\overline{x^2}=0.5$ for sinusoidal modulation.

\section{Training}
\label{sec:training}

The linearised BBFM model shown in Figure \ref{fig:bbfm_linear} was used for training the encoder and decoder parameters using the loss function $L(\mathbf{f},\mathbf{\hat{f})}$ from \cite{valin2024dred}. 
We used a 205 hour speech dataset divided into 4 second sequences for training \cite{valin2024dred}. 
For each 4 second sequence, we chose a random set point received signal level uniformly distributed over a 20~dB range $-100<R_{dBm}<-120$ to encourage operation at a range of SNRs.

\begin{figure}
\begin{center}
\scalebox{.7}{\input {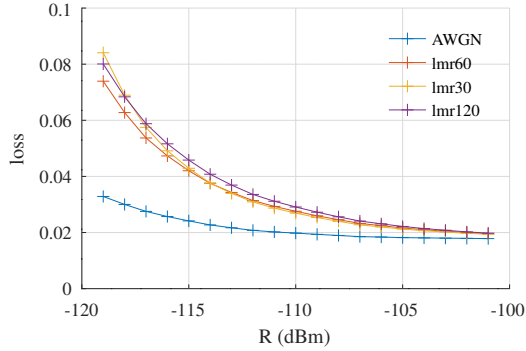}}
\end{center}
\caption{Mean model ``loss" $E[L(\bf{f},\bf{\hat{f}})]$ over the 205 hour training set when tested with fading channels based on different simulated vehicle speeds and an AWGN channel. 
The similar curves for 30, 60 and 120 km/hr indicate the model generalises well.}
\label{fig:loss_R_models}
\end{figure}

The fading model samples were calculated using a worst case fading simulator path delay $d=200$~$ \si{\micro\second}$ and Doppler spreading bandwidth corresponding to a single vehicle velocity of 60 km/hr \cite{telecommunications2016digital}.
We denote this channel \texttt{lmr60}.
Despite training at a single simulated velocity, good results (Figure \ref{fig:loss_R_models}) were obtained when the model was tested at a range of simulated velocities between 30 and 120 km/hr, and on AWGN (non fading) channels.

\section{ASR Evaluation}
\label{sec:asr_results}

\begin{figure}
\begin{center}
\scalebox{1}{\input {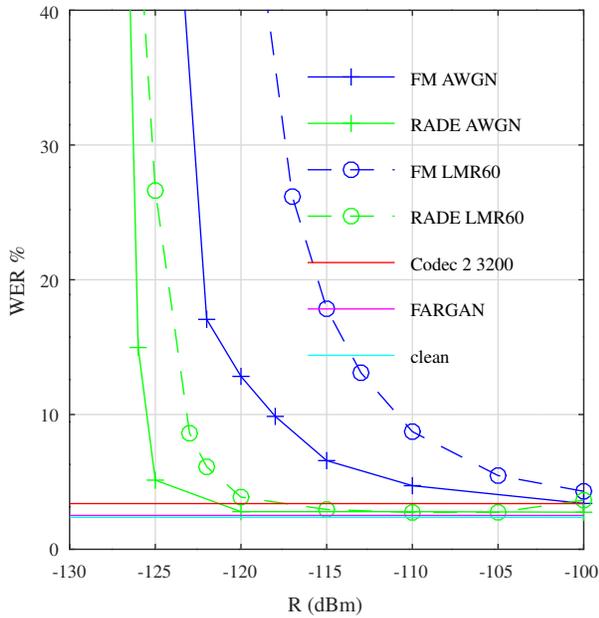}}
\end{center}
\caption{Word error rate \% versus $\mathrm{R_{dBm}}$ for simulated AWGN and \texttt{lmr60} channels, with clean speech, FARGAN with clean features, and Codec 2 at 3200 bit/s with no bit errors as controls.}
\label{fig:wer_r}
\end{figure}

This section describes our evaluation of the RADE-BBFM system and comparison to analog FM.
We have not included a digital LMR protocol as we had difficulty obtaining a full LMR software stack for simulation.
Instead we argue that due to similar speech quality \cite{atkinson2012intelligibility}\cite{vanderau1998delivered}, analog FM is a reasonable proxy for existing digital LMR systems.
We have however included audio samples from a hardware implementation of a digital LMR protocol in our demonstration (Section \ref{sec:demo}). 

Since intelligibility is the primary goal for LMR, we use Automatic Speech Recognition (ASR) to evaluate the performance of the proposed system \cite{rowe2025radeneuralcodectransmitting}.
Five hundred samples from the Librispeech dataset \cite{panayotov2015librispeech} were passed through RADE and analog FM simulations at a range of received signal levels using the AWGN and \texttt{lmr60} channels.
The linearised channel model derived above was used for these simulations.
The output speech was post processed by Whisper ASR system~\cite{radford2022robustspeechrecognitionlargescale}, and the Word Error Rate (WER) measured.
The Librispeech speech and \texttt{lmr60} model datasets used for the evaluation were not part of the training dataset. Hilbert compression is applied to the input speech samples for the analog FM simulation, but not the RADE simulation. 

Fig. \ref{fig:wer_r} presents the results. Compared to analog FM, RADE is very robust to multipath fading.
By drawing lines at constant WER gains of 10~dB can be observed.
The WER of RADE approaches the WER of the baseline FARGAN vocoder at high SNRs.
The Codec 2 vocoder \cite{codec2alg} control simulates the best case performance of classical DSP vocoders employed in current digital LMR protocols.
As expected, it matches the WER of analog FM for strong signals.

\section{UHF Radio Demonstration}
\label{sec:demo}

In this section we describe a demonstration of RADE over UHF radio hardware.
Our demonstration waveform sends 4800 symbols/s over the BBFM channel (a common symbol rate for digital LMR systems), divided into 40~ms frames, or 192 symbols/frame.
The RADE information is sent using 80 symbols/frame, and a 24 symbol unique word is used for frame sync.
The remaining symbols are unused in our system but available for ancillary services such as data and station identification.
Conventional DSP techniques have been used to perform frame sync and timing estimation.

A pair of commercial off the shelf UHF analog FM LMR radios were configured to provide access to the FM modulator and demodulator.
The same radios can be configured to use analog FM to provide a controlled comparison between analog FM and RADE on identical hardware.
AWGN noise is generated by physical processes in the radio receiver front end.
To provide adequate RF isolation the transmit signal is frequency shifted 20~MHz using a mixer and signal generator.
The signal generator drive level to the mixer is used to establish the set point level $R_{dBm}$.
To implement fading the signal generator is amplitude modulated by an arbitrary waveform generator (ARB) driven by calibrated fading magnitude samples generated using Eq. (\ref{eq:fading mag}).

RADE, analog FM, and DStar \cite{dstar_v6} (DStar is a digital LMR system) were tested on AWGN and \texttt{lmr60} channels at a range of received signal levels. In each instance the received levels were confirmed by a calibrated spectrum analyzer.
For each test, a 10 second file containing 4 sentences (2 spoken by male, 2 by a female) was played through the system in real time.

A selection of demonstration audio samples are available on-line at \cite{rade_bbfm}.  Informal listening tests using the demonstration samples show RADE outperforming analog FM and the DStar digital LMR system by a large margin.

\begin{figure}
\begin{center}
\begin{tikzpicture}[auto, node distance=1.5cm,>=triangle 45,x=1.0cm,y=1.0cm,align=center,text width=1cm, font=\footnotesize]

\node [input] (rinput) (z) {};
\node [block, right of=z,node distance=1.25cm] (radio_a) {Radio A};
\node [block, right of=radio_a, text width=0.8cm] (attn1) {-30 dB};
\node [circ, right of=attn1, node distance=1.5cm, text width=0.5cm] (mult) {$\times$};
\node [block, right of=mult, node distance=1.5cm, text width=0.8cm] (attn2) {-30 dB};
\node [block, right of=attn2] (radio_b) {Radio B};
\node [block, below of=mult,node distance=1.5cm] (sig_gen) {Signal Gen};
\node [block, left of=sig_gen, node distance=1.75cm] (arb) {ARB};
\node [output, right of=radio_b,node distance=1.25cm] (routput) {};

\draw [->] node[above,text width=0.5cm] {$\bf{z}$} (rinput) -- (radio_a);
\draw [->] (radio_a) -- (attn1);
\draw [->] (attn1) -- (mult);
\draw [->] (mult) -- (attn2);
\draw [->] (attn2) -- (radio_b);
\draw [->] (radio_b) -- node[above, text width=0.5cm] {$\hat{\bf{z}}$} (routput);
\draw [->] (sig_gen) -- (mult);
\draw [->] (arb) -- (sig_gen);

\node [input,left of=arb,node distance=1cm] (fading) {};
\draw [->] (fading) node[left, text width=0.5cm] {$|H|$} -- (arb); 
\node [input,right of=sig_gen,node distance=1cm] (setpoint) {};
\draw [->] (setpoint) node[right, text width=0.5cm] {$\mathrm{R_{dBm}}$} -- (sig_gen); 

\end{tikzpicture}
\end{center}
\caption{Configuration for UHF radio demonstration.
We inject and sample RADE symbols $\bf{z}$ and $\hat{\bf{z}}$ at the analog FM modulator and demodulator.}
\label{fig:demo_config}
\end{figure}
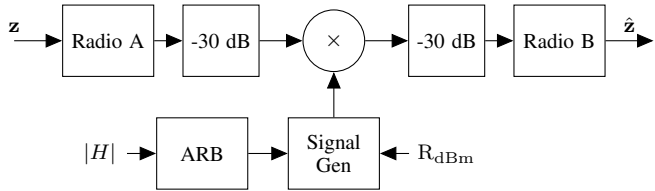

\section{Conclusion}
\label{sec:conclusion}

In this paper we have described a ML based system for speech communication over the ubiquitous, low cost BBFM radio architecture.
Simulations and our demonstration over real radio hardware indicate it outperforms analog FM on AWGN and multipath channels by a significant margin.
This provides a ``drop in" upgrade path for analog FM and classical DSP digital radios based on BBFM hardware.
For the digital radio case the discrete ASK symbols from classical vocoder and FEC components can be replaced by the RADE symbols to provide an upgrade in speech quality and robustness.

The source code for the RADE BBFM system (including channel simulation and model training) is available on GitHub \cite{rade_bbfm_github} in the \texttt{dr-bbfm} branch. 
We would like to acknowledge the contribution of George Karan in conceiving and developing this project and supporting the UHF radio demonstration. 

The complexities and non-linearities in BBFM systems make exact modelling difficult, so several approximations were made to simply training and simulation.
Our simplified linear model does not take into account non-linearities such as the effect of IF filtering on the baseband waveform; we use an approximation of flat versus triangular noise spectra, and a simplified model of the sub-threshold region of FM demodulator operation.
The expressions for FM SNR \cite{crilly2009communication} are only tractable for sine wave modulating signals.
However the significant improvements we have achieved over analog FM indicate these approximations are acceptable, and more accurate models of the BBFM process may not be justified.

The CPU requirements for RADE are high compared to classical vocoders.
As further work investigating optimisations of the RADE encoder, decoder and FARGAN vocoder could significantly reduce this CPU complexity.
However the CPU and memory requirements are already a fraction of that available in common personal communications devices such as mobile (cell) phones, and we anticipate will soon be commonplace in modern LMR radios.

Further work includes research into repeater operation and trunking RADE voice traffic over digital networks.
This would require quantising the continuously valued ASK symbols to a set of discrete levels.
As the waveform is robust to significant AWGN noise we anticipate quantisation noise will not unduly affect performance.

\bibliographystyle{IEEEbib}
\bibliography{2025_rade_bbfm_refs}

\end{document}